\documentclass[]{emulateapj}
\usepackage{graphics}

\newcommand{\gae}{\mathrel{\raise .4ex\hbox{\rlap{$>$}\lower 1.2ex\hbox{$\sim$}
}
}}

\shorttitle{X-ray emission from GB~1508+5714}

\shortauthors{Siemiginowska  et al.}

\begin{document}

\title{An X-ray jet discovered by Chandra
in the z=4.3 radio-selected quasar GB~1508+5714}

\author{Aneta Siemiginowska$^1$, Randall K. Smith$^1$, Thomas
L. Aldcroft$^1$, D.A. Schwartz$^1$\\
Frederic Paerels$^2$ and Andreea O. Petric$^2$ }

\altaffiltext{1}{Harvard-Smithsonian Center for Astrophysics,
Cambridge, MA}
\altaffiltext{2}{Columbia University, New York, NY}

\email{asiemiginowska@cfa.harvard.edu}

\begin{abstract}

We report the {\it Chandra} discovery of an X-ray jet associated with
the redshift 4.3 radio-loud quasar GB~1508+5714. The jet X-ray
emission peaks $\sim$2$\arcsec$ to the South-West of the quasar
core. We present archival HST WFPC2 data of the quasar field which
shows no optical emission at the location of the X-ray jet.  We
discuss possible emission mechanisms and give constraints to the
magnetic field and energy densities for synchrotron radiation or for
Compton scattering of the Cosmic Microwave Background radiation as the
jet X-ray emission process.

\end{abstract}


\section{Introduction}

X-ray jets associated with many quasars observed by the {\it Chandra}
X-ray Observatory are among the recent exciting discoveries in high
energy astrophysics.  The {\it Chandra} data strongly suggest that
jets propagate with high velocities to very large distances from the
quasars (Schwartz et al 2000, Chartas et al 2000, Cellotti et al 2001,
Tavecchio et al 2000, Siemiginowska et al 2002, 2003; Brunetti et al
2002, Sambruna et al 2002).  One possible emission process for these
X-ray jets involves inverse Compton scattering of the Cosmic Microwave
Background (CMB) photons on the relativistic particles within the jet
(Tavecchio et al 2000, Celotti et al 2001). The energy density of the
CMB increases with redshift as $(1+z)^4$, which compensates for the
decrease of surface brightness so that resolved objects with the same
intrinsic properties (particle density, bulk motion, angle to our line
of sight) should be detectable anywhere in the distant universe
(Schwartz 2002a). Detecting X-rays from a sample of high redshift jets
may allow study of the CMB in the early universe.

The highest redshift ($z=2.012$) confirmed X-ray jet published to date
(Fabian et al 2003) is associated with the radio-loud quasar 3C~9.  In
this case the jet X-ray emission is likely due to either Compton
scattered CMB photons (Fabian et al 2003) or thermal emission from gas
heated by jet propagation shocks (Carilli et al 2002).  Schwartz
(2002b) has reported a possible detection of an X-ray jet at the
extreme redshift, $z=5.99$ with no apparent radio counterpart (Petric
et al 2003).  Although this detection needs to be confirmed it hints
at the possibility that at the highest redshifts, X-rays may be the
most efficient wave band to study jets.  Radio-loud quasars at high
redshift are the best candidates for detecting a jet in X-rays,
however, they are quite rare (Snellen et al 2002). There are only 5
redshift z$>$4 radio-loud quasars observed so far with {\it Chandra}.

Here we present a statistically highly significant discovery of an
X-ray jet (123.5$\pm 13.3$ counts) associated with $z=4.3$ radio-loud
flat spectrum quasar GB~1508+5714.  The quasar is X-ray luminous
(L(2-10~keV) = 2.8$\times 10^{46}$ ergs~s$^{-1}$) and Mathur \& Elvis
(1995) and Moran \& Helfand (1996) argue that this luminosity is
partially due to the beaming.  However, the source was not resolved in
radio VLBI observations and there is no detection of a miliarcsec
scale radio jet (Frey et al 1997). Also the published arcsec
resolution VLA radio data do not indicate any structure on the arcsec
scales (Moran \& Helfand 1996).  The peak of the X-ray jet emission is
located at $\sim2\arcsec$ from the quasar core and it is only $\sim
3\%$ of the quasar luminosity.  Detection of the similar radio
emission requires high dynamic range observations not achieved in the
short 5 min exposures.  Here we present the X-ray data of the quasar
and the jet, and discuss the possible jet emission mechanisms.

Throughout this paper we use the cosmological parameters based on the
WMAP measurements (Spergel et al. 2003):
H$_0=$71~km~sec$^{-1}$~Mpc$^{-1}$, $\Omega_M = 0.27$, and $\Omega_{\rm
vac} = 0.73$. At $z= 4.3$, 1~arcsec corresponds to 6.871~kpc.

\section{Chandra Observations}

Q1508+5714 \ was observed with the {\it Chandra} Advanced CCD Imaging
Spectrometer (ACIS-S, Weisskopf et al 2002) on 2001 June 13 (ObsID
2241).  The source was located on the S3 chip (BI) and offset by $\sim
35\arcsec$ from the optical axes. The observation was made in the FAINT
mode with the 3.241~sec frame readout time of the full CCD.  The
source count rate of ~0.054 cts/sec is relatively low and the
observation is not affected by pileup. At this count rate we expect
only $\sim$5 counts in the readout streak.  Note that this {\it
Chandra} observation was first presented by Telis et al (2002) in the
context of X-ray dust halos.

We have reprocessed the archival data using the {\it Chandra} CALDB
version 2.22. We ran {\tt acis\_process\_events} to remove pixel
randomization to obtain the highest image resolution data. The X-ray
position of the quasar (J2000: 15 10 02.89, +57 02 43.32) agrees with
the radio position (Ma et al. 1998) to better than 0.05$\arcsec$,
(which is smaller than {\em{Chandra}}'s 90$\%$ pointing accuracy
of 0.6~arcsec\footnote{http://cxc.harvard.edu/cal/ASPECT/celmon/}), so
we have high confidence in the source identification.  We used CIAO
3.0
\footnote{http://cxc.harvard.edu/ciao/} software to analyze the data.
The standard data filtering leaves the total exposure time at
$\sim$88.97~ksec.

\subsection{Image Analysis}
\label{image}

\begin{figure}
\resizebox{2.8in}{!}{\includegraphics{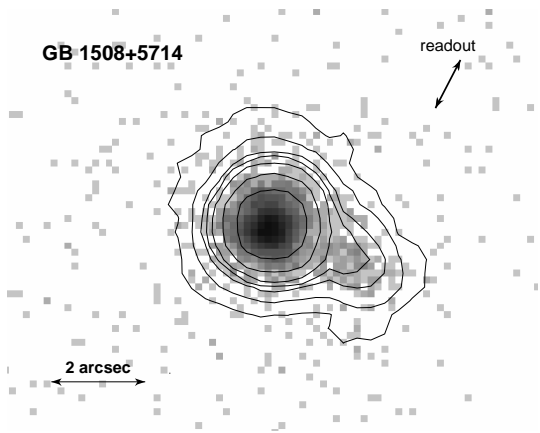}}
\caption{The observed {\it Chandra} ACIS-S image of
GB~1508+5714.  The spatial scale is indicated by a 2 arcsec arrow.
The readout direction is indicated by arrows in the upper right
corner.  The North is up and the East is left. The pixel size
corresponds to 0.148$\arcsec$. The image is in logarithmic scale and
contour levels are at 0.2, 0.5, 1.2, 1.8, 3.0 ,7.3, 23
counts/pixel. The maximum of 198 counts per pixel in the quasar core.}
\label{fig:acis}
\end{figure}

The binned ACIS-S data are displayed in Fig.~\ref{fig:acis}.  The
source has a clear extension to the South-West (PA $\sim-114^{o}$)
which is inconsistent with the HRMA (High Resolution Mirror Assembly)
point spread function (PSF).  

\begin{figure}
\centerline{
\resizebox{2.8in}{!}{\includegraphics{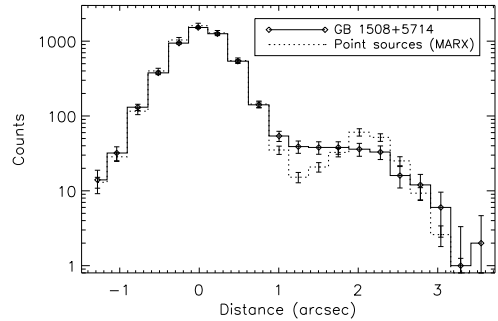}}}
\caption{Profile along the direction of extended emission for Chandra
observation of GB~1508+5714 (solid line) and for a simulated
observation using two point sources separated by 2\arcsec\ (dotted
line).  The central core emission matches well but the extended
emission (from 1\arcsec\ to 3\arcsec\ from core) is highly
inconsistent with a point source.}
\label{fig:sim}
\end{figure}

To determine if the extended emission could be due to a nearby point
source (e.g. a lensed image or binary quasar), we ran a ray trace
using CHART and then
MARX\footnote{http://cxc.harvard.edu/chart/} to create a high S/N
simulation of two point sources separated by 2.0\arcsec.  (The
centroid of the extended emission is $\sim$2.0\arcsec\ from the quasar
core.)  We modeled both the quasar core and the extended emission as
point sources with energy spectra given by the fitting described in
Section~\ref{qso}. We then extracted a linear profile along the
direction of the extended emission, using 0.25\arcsec\ slices oriented
perpendicular to the direction of extent.  For the simulation we added
7$\%$ errors to account for uncertainty in the raytrace model.  The
resultant profiles are shown in Fig~\ref{fig:sim} and clearly
illustrate that the observed extended emission (solid line) between
1\arcsec\ to 3\arcsec\ is inconsistent with a point source (dashed
line).  Using $\chi^2$ statistics, the probability that the profiles
are consistent over that distance range is less than $10^{-10}$.

In Section~\ref{sec:hst} we present an HST image which rules out
gravitational lensing or a foreground galaxy. We conclude that the
quasar image is extended and that the South-West feature has a linear
extent which is most plausibly due to an X-ray jet.  The chance
probability of finding a source at the jet flux level within
10$\arcsec$ of the quasar is very low, approximately 0.5$\%$.  This is
based on the Chandra Deep Field observations (Giacconi et al, 2001),
which give 200~sources~deg$^{-2}$ with flux above 5$\times
10^{-15}$ergs~cm$^{-2}$sec$^{-1}$.

\vspace*{2em}

\subsection{Spectral Modeling of the Quasar}
\label{qso}

\begin{figure}
\resizebox{2.8in}{!}{\includegraphics{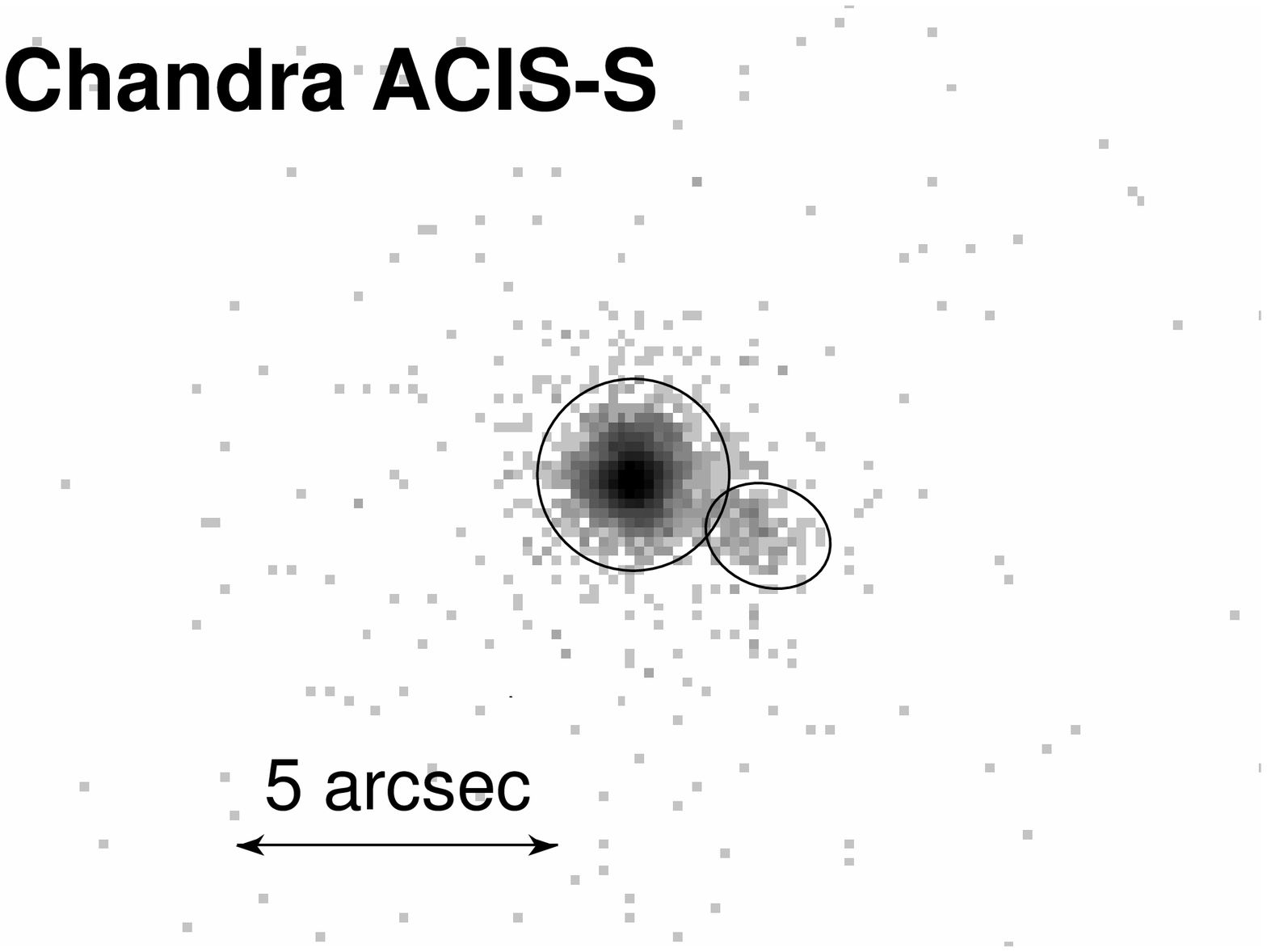}}
\caption{ The ACIS-S data shows the quasar and the extension into 
the South-West. The quasar and jet spectrum extraction regions are
overlayed on the image. The pixel size corresponds to 0.148
arcsec. The spatial scale is indicated with the arrow. The image is in
logarithmic scale with the maximum of 198 counts within the core
pixel. The North is up and the East is left.}
\label{fig:acis-regions}
\end{figure}

The quasar spectrum with a total of 5242 counts was extracted from a
1.5\arcsec\ circle centered on the pixel (x,y)=(4022.06,4118.96)
(physical coordinates, see Fig.\ref{fig:acis-regions}).  Based on the
CHART simulations (Sec.~\ref{image}) we estimate that $\sim$95$\%$ of
the source counts lie within this source region.  
The background (annulus between 7.5$\arcsec$ and 15$\arcsec$)
intensity is low with only 8.8$\pm$0.4 counts expected in the
source regions.  The energy response of ACIS-S below 1~keV is affected
by a contamination
layer\footnote{http://cxc.harvard.edu/ciao/caveats/}.  To account for
the contaminant we have applied a correction using {\tt apply\_acisabs
v.1.1-2}\footnote{http://cxc.harvard.edu/ciao/threads/sherpa\_acisabs/}.
Still there are uncertainties of order $\sim 15\%$ (for energies below
1~keV) due to unknown properties of the
contaminant\footnote{http://cxc.harvard.edu/cal/}.

We used {\it Sherpa} (Freeman et al 2001) to fit the quasar spectrum
(0.3-10.0~keV) with an absorbed power law model assuming Galactic
equivalent hydrogen column of N$^{gal}_H$= 1.46$\times 10^{20}$
atoms~cm$^{-2}$ (Stark et al 1992).  The best fit parameter values are
listed in Table~\ref{tab:xray-fits}.  The observed 2-10 keV quasar
flux based on this model is equal to
3.45$\times 10^{-13}$ergs~cm$^{-2}$~sec$^{-1}$ 
The rest frame 2-10~keV quasar luminosity is equal to 2.84$\times
10^{46}$ ergs~cm$^{-2}$, while the total luminosity observed in
0.2-10~keV (rest frame 1.6-53~keV) is equal to 2.32$\times 10^{47}$
ergs~cm$^{-2}$.

The observed photon index $\Gamma = 1.55\pm 0.05$ is consistent with
the first ASCA observation from March 1995 (Moran \& Helfand 1997)
while the normalization of the power law in the ASCA observation is
twice that obtained with ACIS-S. This agrees with the variability of
the source as claimed by Moran \& Helfand.

\subsection{Spectral Modeling of the X-ray Jet}

We extracted the jet ACIS-S spectrum from the elliptical region
centered on the pixel \break (x,y)=(4026.075,4117.125) (J2000: 15 10
02.648, +57 02 42.39) as shown in Fig.~\ref{fig:acis-regions}. The
total of 149 counts from that region were binned to have a minimum of
10 counts per energy bin. We assume the background spectrum from an
annulus between 1.3$\arcsec$ and 3.3$\arcsec$ excluding an elliptical
jet region.  There are 123.5$\pm 13.3$ net counts detected in the jet
with a maximum energy of $\sim$5~keV ($\sim$26.5 keV in the rest
frame).

We fit the spectrum between 0.3-7.0~keV with an absorbed power law
model assuming the total N$_H$ column of 2.7$\times 10^{20}$cm$^{-2}$
as determined by fitting the quasar spectrum.
Table~\ref{tab:xray-fits} lists the best fit model parameters. The jet
photon index $\Gamma = 1.90\pm 0.36$ is steeper than the quasar one.
The model gives an observed 2-10 keV flux of 4.84$\times
10^{-15}$ ergs~cm$^{-2}$~sec$^{-1}$ and 3.86$\times
10^{-15}$ergs~cm$^{-2}$~sec$^{-1}$ (absorbed) within 0.1-2 keV.
Assuming isotropic emission we compute the jet 2-10~keV (rest frame)
luminosity of 7$\times 10^{44}$ ergs~sec$^{-1}$. This luminosity is
similar to the luminosity of the X-ray jets at lower redshifts. The
ratio of the quasar and the jet X-ray luminosities is similar to the
one found in lower redshift objects (Schwartz et al 2000, Sambruna et
al 2002, Siemiginowska et al 2002).

\subsection{HST Image.}
\label{sec:hst}

\begin{figure}
\resizebox{2.8in}{!}{\includegraphics{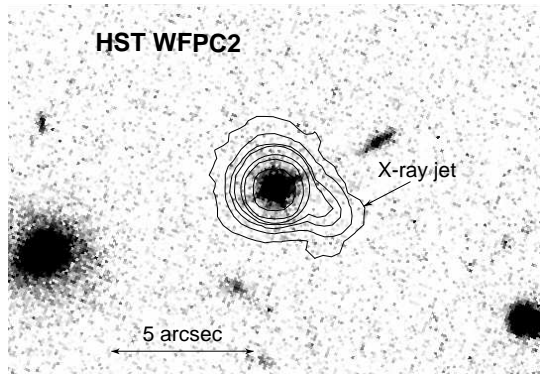}}
\caption{HST WFPC2 image of the quasar field with the X-ray contours
overlayed. No source is present in the jet region.  The scale is
indicated with the arrow. The North is up and the East is left.}
\label{fig:hst}
\end{figure}

GB~1508+5718 was observed with HST/WFPC2 (F814W filter) for 
4800~s (four exposures) on 22-July-1995.  We retrieved the single
combined image and variance image created by the WFPC2 Associations
Pipeline\footnote{http://archive.stsci.edu/hst/wfpc2/about.html}.
Fig.~\ref{fig:hst} shows the X-ray contours overlayed on top of the
HST/WFPC2 image. Using aperture photometry with the same quasar and
jet regions given in the X-ray analysis section (also
Fig.~\ref{fig:acis-regions}), we derive the I-band magnitudes $I_{\it
quasar} = 19.2$ and $I_{\it jet} > 25.3$ (3-$\sigma$ limit).


The large difference in brightness between the quasar and the optical
limit to the jet emission excludes the possibility that the X-ray
emission is due to gravitational lensing. The flux ratio between the
quasar and the jet is equal to 40 in the X-rays, while it is more than
$\sim$293 in the optical.

\section{Discussion}

X-ray emission from jets is due to either the synchrotron process or
Compton scattering of seed photons (synchrotron, SSC or from outside
the jet, IC) off the relativistic particles in the jet (see Harris \&
Krawczynski 2002 for a review).  The IC scenario with CMB radiation as
a source of the external photons was proposed by Tavecchio et al
(2000) and Celotti et al (2001) for jets where the SSC model predicts
an X-ray flux which is too low to match the data. This model requires
the jet to move with a Lorentz factor, $\Gamma_{bulk}$, of
$\sim$3-10. Jet electrons with relatively low energy ($\gamma
\sim$ 100-1000) can then Compton scatter the CMB photons
into the X-ray band. Note that the synchrotron emission from these low
energy electrons may not be detectable in radio because it will be
emitted at very low frequencies.

Which process dominates the X-ray jet emission in GB~1508+5714?  The
HST optical limit is consistent with both synchrotron or Compton
scattering processes.  The X-ray jet is detected up to $\sim$26~keV
($\nu _{max} \sim 6.5\times 10^{18}$~Hz) in the quasar frame and we
cannot constrain the high energy turn-over in the X-ray spectrum.  If
the break occurs at $\sim$26~keV then the synchrotron emitting
electrons have energies $\gamma \sim 10^9$.  The lifetime of such
electrons is short ($ < 10$ years for an equipartition magnetic field)
implying that they need to be accelerated very recently and in highly
efficient process if the X-rays are due to the synchrotron emission.

There is no reported detection of the radio jet (Frey et al 1997) and
the VLA observations are consistent with 95$\%$ of the 5~GHz flux
being emitted by the quasar (Moran \& Helfand 1997).  Assuming that
the X-ray emission is due to synchrotron process we can extrapolate
the X-ray spectrum into the radio band.  For the observed photon index
$\Gamma = 1.9$ and 1~keV flux density of 1.68$\times
10^{-6}$~photons~cm$^{-2}$~sec$^{-1}$~keV$^{-1}$ we estimate the 5~GHz
flux density to be $\sim$9~mJy.  This is about $6\%$ of the radio flux
from the quasar at this frequency, and is consistent with no radio
detection.

The projected size of the jet corresponds to $\sim$15~kpc.  The high
luminosity of the quasar core may be due to beaming (Mathur
\& Elvis 1995), so the viewing angle of the jet might be small and 
therefore the jet could be much longer (e.g. $\sim150$~kpc for
$\theta\sim6\deg$). Such scales compare well with lower redshift X-ray
jets where the IC/CMB process may dominate the jet X-ray emission
(Siemiginowska et al 2002, Sambruna et al 2002).  The energy density
of the CMB at redshift z=4.3 is 3.3$\times 10^{-10}$~ergs~cm$^{-3}$
(for the CMB radiation temperature at z=0 of 2.728~K,
Fixen et al 1996). At the quasar redshift the CMB energy density will
dominate magnetic fields of less than 91$\mu G$.

We can calculate the equipartition magnetic field assuming the radio
flux upper limit of 9 mJy at 5 GHz, and a uniform volume distribution
of fields and particles filling a cylindrical region (1.6$\arcsec$
long and 0.4$\arcsec$ in radius, volume of $\sim$7.7$\times 10^{66}$
cm$^3$).  This gives B$_{eq} \sim$ 268~$\mu$G.
However, if the same electron population also
produces the observed X-ray flux via IC/CMB, then from Felten \&
Morrison (1966) we calculate B$_{IC} \sim$ 25~$\mu$G.  These two
values can be reconciled when we consider a relativistic jet with an
effective Doppler factor $\delta$ (Tavecchio et al. 2000, Celotti et
al. 2001)
because B$_{eq}$ $\propto 1/\delta$ while B$_{IC}$ $\propto
\delta$, so 
we can find a self-consistent solution 
at B=83~$\mu$G and $\delta$=3.2.  Because we used an upper limit to
the radio flux our number for the magnetic field is also an upper
limit, while the value of $\delta$ is a lower limit. These values have
an uncertainty due to the uncertain X-ray slope (the assumed radio
spectral index) giving the parameters range of ($B$=161$\mu$G,
$\delta$=2.6) and (B=37$\mu$G, $\delta=5.3$).


From the upper limit to the radio luminosity and the magnetic field
given above we estimate an upper limit on the photon density of the
synchrotron radiation of $\sim 2 \times 10^{-12}$ ergs~cm$^{-3}$ (for
the cylindrical region)
This is smaller than the energy density of the magnetic field and the
CMB radiation.  The energy density of the CMB radiation in the jet's
comoving frame is higher by a factor of $\Gamma_{bulk}^2$ and even for
moderate jet velocities ($\Gamma_{bulk}
\ge 1$): $u'_{CMB}=3.3\times 10^{-10} (\Gamma_{bulk}^2 - 0.25)$
(Harris \& Krawczynski 2001) which dominates synchrotron radiation
field.  The SSC emission will be too weak in comparison with the
IC/CMB to dominate the X-ray jet spectrum.

It is quite likely that the X-ray emission in the GB~1508+5714 jet is
due to the interaction between the CMB photons and the relativistic
jet particles. High quality radio data are necessary to constrain the
model and exclude the synchrotron possibility.

A sample of high-redshift X-ray jets may provide a way to study the
evolution of the CMB with redshift, which is a fundamental prediction of
standard Big Bang cosmology.  Only recently has the non-local CMB
temperature been measured (using quasars absorption lines) and shown to be
higher at redshift 2.33 than at z=0 (Srianand, Petitjean \& Ledoux 2000).  
Schwartz (2002a) argues that there should be many IC/CMB dominated X-ray
jets at high redshift.  A sample of such sources would allow us to
estimate the CMB intensity as a function of redshift.\\



\acknowledgements

We wish to thank the anonymous referee for helpful comments which
substantially improved the paper. We thank Martin Elvis for
discussions.  This research is funded in part by NASA contracts
NAS8-39073 to the {\it Chandra} X-ray Center, and
part by NASA through Chandra Award Number GO2-3148A issued by
the Chandra X-Ray Observatory Center, which is operated by the
Smithsonian Astrophysical Observatory for and on behalf of NASA under
contract NAS8-39073.


\begin{table}[h]
\begin{scriptsize}
\begin{center}
\caption{Model Parameters}
\medskip
\begin{tabular}{lccccccc}
\hline \hline \\
 Model & $z_{abs}$$^a$ &  N$_H$($z_{abs}$)$^b$ & $\Gamma$ & Norm$^c$ & $\chi ^2$(d.o.f.)$^d$ \\
\hline
GB~1508+5714      &  0        & 0.13$\pm 0.1$ & 1.55$\pm0.06$ & 6.76$\pm0.33$ &
540.4 (660) \\
GB~1508+5714     &  4.3        & 5.5 $\pm 3.4$ & 1.56$\pm0.05$ &
6.73$\pm0.25$ & 538.4 (660) \\
Jet      &  0     & 0.27$^e$  & 1.90$\pm0.36$ & 0.169$\pm0.03$ &
27.8 (42) \\
\hline
\end{tabular}

\end{center}

Model: $N(E)= Norm \, E^{-\Gamma}*\rm exp[- N^{gal}_H \sigma (E) - N^{z_{abs}}_H
\sigma(E(1+z_{abs}))]$~photons~cm$^{-2}$~sec$^{-1}$~keV$^{-1}$;
N$^{gal}_H$ = 1.4$\times 10^{20}$atoms~cm$^{-2}$; $\sigma (E)$ and
$\sigma E(1+z_{abs})$ - absorption cross sections (Morrison \&
McCammon 1983, Wilms, Allen \& McCray 2000).  All errors are 90$\%$;
$^a$ redshift of the absorber; $^b$ equivalent Hydrogen absorbing
column in excess to the Galactic column in units of 10$^{21}$
atoms~cm$^{-2}$; $^c$ Normalization of the power law at 1keV in units
of 10$^{-5}$photons~cm$^{-2}$~s$^{-1}$; $^d$ $\chi^2$ calculated with
Primini method using {\it Sherpa}, degrees of freedom are given in the
bracket.$^e$ the assumed column is equal to a sum of Galactic column
and the excess column N$_H(z_{abs}=0)$ from the fit to the quasar
spectrum.

\label{tab:xray-fits}
\end{scriptsize}

\end{table}






\end{document}